# Observation of universal strong orbital-dependent correlation effects in iron chalcogenides

M. Yi[1,2†], Z.-K. Liu[1,2†], Y. Zhang[1,3†], R. Yu[4,5], J.-X. Zhu[6], J. J. Lee[1,2], R. G. Moore[1], F. T. Schmitt[1], W. Li[1], S. C. Riggs[1,2], J.-H. Chu[7], B. Lv[8], J. Hu[9], M. Hashimoto[10], S.-K. Mo[3], Z. Hussain[3], Z. Q. Mao[9], C. W. Chu[8], I. R. Fisher[1,2], Q. Si[5], Z.-X. Shen[1,2*], and D. H. Lu[10*]

[1]*Stanford Institute for Materials and Energy Sciences, SLAC National Accelerator Laboratory and Stanford University, Menlo Park, California 94025, USA*

[2]*Departments of Physics and Applied Physics, and Geballe Laboratory for Advanced Materials, Stanford University, Stanford, California 94305, USA*

[3]*Advanced Light Source, Lawrence Berkeley National Lab, Berkeley, California 94720, USA*

[4]*Department of Physics, Renmin University of China, Beijing 100872, China*

[5]*Department of Physics and Astronomy, Rice University, Houston, Texas 77005, USA*

[6]*Theoretical Division, Los Alamos National Laboratory, Los Alamos, New Mexico 87545, USA*

[7]*Department of Physics, University of California, Berkeley, California 94720, USA*

[8]*Department of Physics, Texas Center for Superconductivity, University of Houston, Houston, Texas 77204, USA*

[9]*Department of Physics and Engineering Physics, Tulane University, New Orleans, Louisianna 70118, USA*

[10]*Stanford Synchrotron Radiation Lightsource, SLAC National Accelerator Laboratory, Menlo Park, California 94025, USA*

[†]*These authors contributed equally.*

*To whom correspondence should be addressed: dhlu@slac.stanford.edu and zxshen@stanford.edu

**Establishing the appropriate theoretical framework for unconventional superconductivity in the iron-based materials requires correct understanding of both the electron correlation strength and the role of Fermi surfaces. This fundamental issue becomes especially relevant with the discovery of the iron chalcogenide superconductors. Here, we use angle-resolved photoemission spectroscopy to measure three representative iron chalcogenides, $FeTe_{0.56}Se_{0.44}$, monolayer FeSe grown on $SrTiO_3$, and $K_{0.76}Fe_{1.72}Se_2$. We show that these superconductors are all strongly correlated, with an orbital-selective strong renormalization in the $d_{xy}$ bands despite having drastically different Fermi surface topologies. Furthermore, raising temperature brings all three compounds from a metallic state to a phase where the $d_{xy}$ orbital loses all spectral weight while other orbitals remain itinerant. These observations establish that iron chalcogenides display universal orbital-selective strong correlations that are insensitive to the Fermi surface topology, and are close to an orbital-selective Mott phase, hence placing strong constraints for theoretical understanding of iron-based superconductors.**

## Introduction

Since the discovery of high temperature superconductivity in the iron pnictides (FePn), extensive research efforts have revealed many common properties of these materials. In the vast material base of the FePns, all parent phases are metallic, and the observed electronic structures are largely consistent with the prediction of *ab initio* LDA calculations[1-2]. These properties have led to theoretical understanding that the electron correlations in FePns are much weaker than in cuprate superconductors, whose parent phases are Mott insulators. Furthermore, the observation of comparable sized hole pockets at the Brillouin zone (BZ) centre and electron pockets at the BZ corner have resulted in the proposal that such a FS topology is ubiquitous and essential to superconductivity in iron pnictides, and pairing in the iron pnictides is mediated by antiferromagnetic fluctuations via Fermi surface (FS) nesting between the hole and electron Fermi pockets[3]. On the other hand, there have also been theoretical proposals of 'incipient Mott localization' for which the system is metallic but on the verge of localization[4-6], supported by the normal state bad metal behaviour from optical conductivity measurements[7] and the large fluctuating magnetic moment comparable to the antiferromagnetic insulating copper oxides[8]. This discussion became especially relevant with the more recent discovery of iron chalcogenides (FeChs)[9-11], which not only possess large local magnetic moments[12] and insulating phases[13], but also include compounds that lack hole pockets needed for FS nesting yet have comparable $T_C$'s as FePns[14-17].

There are currently three major classes of FeCh superconductors: $FeTe_{1-x}Se_x$, $K_xFe_{2-y}Se_2$, and FeSe film grown on $SrTiO_3$. In the $FeTe_{1-x}Se_x$ family, superconductivity is achieved with isovalent substitution of Se for Te that suppresses magnetic order in the FeTe end[18], and becomes optimal in $FeTe_{0.56}Se_{0.44}$ (FTS), where $T_C$ is 14.5 K. The $K_xFe_{2-y}Se_2$ family has in its phase diagram insulating phases with magnetic moments as large as $3.3\mu_B/Fe$[10,12-13], and optimal $T_C$ is 32 K, such as achieved in $K_{0.76}Fe_{1.72}Se_2$ (KFS). FeSe film grown on $SrTiO_3$ is the latest addition to the FeCh family, with monolayer film (FS/STO) having a record $T_C$ possibly exceeding 65 K[11,16-17]. In a previous angle-resolved photoemission spectroscopy (ARPES) study[19] on KFS, we found the low temperature state to be a metallic state with orbital-dependent renormalization— where the $d_{xy}$ orbital dominated bands are strongly renormalized as compared to other orbitals. Raising temperature drives the material to an orbital-selective Mott phase (OSMP) in which the $d_{xy}$ orbital completely loses spectral weight while other orbitals remain itinerant. Subsequently, such a temperature scale was also identified by THz spectroscopy[20], Hall measurements[21] and pump-probe spectroscopy[22], where the slight temperature scale variations are due to the different definitions used. This motivates the usage of orbital selectivity to address the fundamental open question of whether it is the electron correlation strength or the nature of the FS that plays a predominant role in the microscopic physics of the FeCh materials.

In this work, we study systematically the optimal superconducting members of the three FeCh families, FTS, KFS, and FS/STO, using ARPES. We show that, in the low temperature state, in contrast to the FePns, all of the FeChs are in a strongly correlated regime where strong orbital-selective renormalization is observed on the $d_{xy}$ bands, despite having drastically different FS topologies. In addition, by increasing temperature, all of the FeChs crossover into a phase where the $d_{xy}$ orbital completely loses spectral weight while other orbitals remain metallic. These observations showcase the universally strong orbital-selective electron correlations in the FeChs, and that the superconductivity in the FeChs, which is insensitive to FS topology, occurs in proximity to an orbital-selective Mott phase, placing strong constraints on the theoretical understanding of the iron-based superconductors.

**Results**

*Orbital-selective band renormalization at low temperatures*

The generic electronic structure of iron-based superconductors (FeSC) consists of three hole bands at the BZ centre, $\Gamma$, and two electron bands at the BZ corner, $M$. The hole bands are predominantly of $d_{xz}$, $d_{yz}$, and $d_{xy}$ orbital characters, while the electron bands are $d_{xz}$ and $d_{xy}$ along $\Gamma$-$M$. The relative positions of these bands with respect to each other as well as to the Fermi level ($E_F$) could vary with differences in lattice parameters and doping level. Hence, the Fermi surface topology among different FeSCs could be qualitatively different, as shown in Fig. 1a-d, where the Fermi pockets at the BZ centre vary from being hole-like to non-existent to electron-like. The measured band structure along the $\Gamma$-$M$ high symmetry direction for the three compounds are shown in Fig. 1e-g, in comparison to that for the optimally Co-doped BaFe$_2$As$_2$ (BFCA) (Fig. 1h), an iron pnictide as a reference. For FTS (Fig. 1i), one of the hole bands crosses $E_F$, and both electron bands cross $E_F$ at $M$, resulting in roughly compensated hole pocket at $\Gamma$ and electron pockets at $M$ (Fig. 1a), consistent with isovalent substitution for this compound. For both FS/STO (Fig. 1j) and,KFS (Fig. 1k) in contrast, only the electron bands cross $E_F$ while the hole band tops are well below $E_F$, with an additional small electron pocket at $\Gamma$ in KFS. Thus, there is heavy electron doping in both compounds as reflected in a Fermi surface topology consisting only of electron pockets (Fig. 1b-c). Comparing the band structure of the three FeChs to the FePns, we notice a significant difference near the $M$ point—there is an apparent gap between the bottom of the electron bands and the top of the hole band in all three FeChs, in sharp contrast to BFCA (Fig. 1l), in which the d$_{xz}$ electron band bottom is degenerate with the d$_{yz}$ hole band top.

Generally in FeSC, this degeneracy between the d$_{xz}$ electron band bottom and d$_{yz}$ hole band top at the zone corner is protected by the C$_4$ rotational symmetry, as seen in BFCA (Fig. 1l) and corresponding LDA calculations (Fig. 2a). This degeneracy is only lifted with the breaking of C$_4$ symmetry, as in the orthorhombic phase of underdoped BFCA[23], NaFeAs[24-25], bulk FeSe[26-29], and multilayer FeSe film[30-31], where a splitting between the corresponding d$_{xz}$ bands and d$_{yz}$ bands occurs, in addition to a doubling of the bands from twinning effects due to the orthorhombic distortion. However, no static C$_4$ symmetry breaking has been reported for any of the FeChs studied here, nor are twinning effects observed here that is expected from a broken symmetry due to orthorhombic distortion. Rather, this apparent gap can be explained by a strong orbital-dependent band renormalization and band hybridization. As the schematic shown in Fig. 2, the LDA calculated $d_{xy}$ electron band bottom is deeper than that of the $d_{xz}$ band. If the $d_{xy}$ orbital is strongly renormalized compared to the other orbitals, the $d_{xy}$ electron band bottom, i.e., the corresponding $d_{xy}$ hole band top, would rise above that of the $d_{xz}$ electron band (Fig. 2b). The heavily renormalized $d_{xy}$ hole band then crosses the $d_{xz}$ electron band and the two bands hybridize such that a gap appears at the $M$ point without lifting the $d_{xz}/d_{yz}$ degeneracy protected by $C_4$ symmetry (Fig. 2c). Evidence for two nearly degenerate electron bands can be seen in the high resolution spectra acquired on FS/STO (Fig. 3a). As an aside, we note that in the unrenormalized case (Fig. 2a), a hybridization gap between the $d_{xy}$ electron band and $d_{yz}$ hole band is not observed. This is because when considering the hopping via the chalcogen atoms along this high symmetry direction ($x$), both $d_{xy}$ and $d_{xz}$ are odd while $d_{yz}$ is even. Hence $d_{xy}$ does not mix with $d_{yz}$ to produce a hybridization gap[32].

This interpretation is further supported by the observed strong renormalization of the $d_{xy}$ hole band near $\Gamma$, which is significantly more renormalized than the $d_{xz}/d_{yz}$ hole bands, clearly seen in all three compounds (Fig. 1i-k). For FTS, the $d_{xy}$ hole band is strongly renormalized by a factor of ~16 compared to LDA calculations, while $d_{xz}$ and $d_{yz}$ bands are only renormalized by factors of ~4. Moreover, in the FeTe$_{1-x}$Se$_x$ family, it has been found that the $d_{xy}$ band renormalization factor strongly increases towards the FeTe end compared to that of $d_{xz}/d_{yz}$, further revealing the strong orbital-dependence in the itinerant to localized crossover in this system[33]. For KFS, the $d_{xy}$ hole band is renormalized by a factor of ~10 compared to the factor of ~3 for $d_{xz}/d_{yz}$ bands[19]. This is even more apparent in FS/STO, where enhanced $d_{xy}$ orbital matrix elements in the second BZ shows the nearly flat $d_{xy}$ hole band extending towards the bottom of the shallow electron bands at $M$ with a noticeable hybridization gap (Fig. 3b). In contrast, in BFCA, the renormalization factor for $d_{xy}$ hole band is comparable to that of the $d_{xz}/d_{yz}$ hole bands $(2\sim3)^2$, as can be seen in the band slopes in the second derivative plot (Fig. 1l). In summary, all three FeCh systems show much stronger renormalization in the $d_{xy}$ orbital compared to the $d_{xz}/d_{yz}$ orbitals in the low temperature state, in contrast to the FePns. Here we would like to note that while ARPES is a surface-sensitive probe, the universal orbital-selective renormalization among the FeChs despite their different structure and surface terminations indicate that these properties must represent the bulk, rather than a result of extrinsic variation of surface properties

*Temperature dependence*

Next, by raising temperature sufficiently high, we notice that in all three FeCh systems, the $d_{xy}$ orbital-dominated bands lose spectral weight completely (Fig. 4), as reported previously for KFS[19]. This can be seen first in the disappearance of the shallow $d_{xy}$ hole band near the $\Gamma$ point. In the low temperature state of FTS, the $d_{xy}$ hole band crosses the $d_{yz}$ hole band near $\Gamma$, and a small hybridization gap appears as can be seen in the discontinuous intensity pattern of the strong $d_{yz}$ hole band (Fig. 4a). At high temperature, this discontinuity disappears as only the $d_{xz}$ and $d_{yz}$ hole bands remain (Fig. 4d). In KFS, the nearly flat $d_{xy}$ hole band in the low temperature state is entirely above the $d_{xz}/d_{yz}$ hole band tops (Fig. 4c), and clearly disappears at high temperatures (Fig. 4f). The second evidence for the disappearance of $d_{xy}$ orbital at high temperatures is the vanishing of the apparent gap between the electron band bottom and the $d_{yz}$ hole band top at $M$. As shown in the schematic (Fig. 4g-h), when the $d_{xy}$ band disappears, its hybridization gap with the $d_{xz}$ electron band naturally vanishes, recovering the deeper non-hybridized $d_{xz}$ electron band whose bottom is degenerate with the $d_{yz}$ hole band top as expected. This is clearly seen in the high temperature data in all three systems (Fig. 4d-f). Here we note that while KFS is known to have phase separation issues with the existence of superconducting regions and insulating regions[19], FTS and FS/STO, on the other hand, do not have such complication. Hence the universal behaviour of vanishing $d_{xy}$ spectral weight with raised temperature observed here reflects an intrinsic effect rather than a result of intricate phase separation in KFS.

To examine this temperature dependence more carefully, we have quantitatively analysed the spectral weight of each FeCh system. For FTS, we track the $d_{xy}$ hole band slightly away from the $\Gamma$ point where it is well separated from the $d_{yz}$ hole band (Fig. 5a). From the energy distribution curve (EDC) at this momentum, we fit Gaussian peaks for both $d_{xy}$ hole band near $E_F$ and $d_{yz}$ hole band at higher energies, along with a Shirley background (Fig. 5d), and track the integrated spectral weight of the $d_{xy}$ and $d_{yz}$ peaks as a function of temperature. Comparing these two orbitals, we see that the $d_{xy}$ spectral weight drops to zero around 110 K, in contrast to a very weak decrease of the $d_{yz}$ spectral weight. This is very similar to the situation in FS/STO and KFS, where we track the spectral weight of the $d_{xy}$ orbital at the $d_{xy}$ electron band bottom at $M$. For KFS, the fitted peak area precipitously drops around 100 K, and approaches zero above 180 K (Fig. 5i). For FS/STO, the $d_{xy}$ spectral weight approaches zero above 150 K while that of the $d_{yz}$ orbital remains finite (Fig. 5h), demonstrating the orbital-dependence of this temperature evolution.

*Theoretical calculations*

For all three FeCh superconductors studied, two observations are universal: i) strong orbital-dependent renormalization at low temperatures, ii) disappearance of $d_{xy}$ spectral weight with raised temperatures. Theoretically, these behaviours can be understood in proximity to an OSMP, as predicted by both a slave-spin mean-field method[34] and dynamical mean field theory method[35] taking into account sufficiently strong intra-orbital Coulomb repulsion $U$ and Hund's coupling $J$. These works show that the FeChs are close to an OSMP in which the $d_{xy}$ orbital is completely Mott localized while the other orbitals remain itinerant, and that a temperature scale could be identified where the $d_{xy}$ orbital loses coherence and spectral weight, together with its hybridization to the other orbitals.

The orbital-selectivity arises from two factors[19,34]. The crystal field splitting of the tetragonal lattice makes the $d_{xy}$ orbital energy level higher than that of $d_{xz}/d_{yz}$. Because the threshold interaction for the Mott transition is larger for degenerate orbitals than for non-degenerate orbitals[36], it is easier to localize the $d_{xy}$ electrons. The tendency towards OSMP is further enhanced when the projected bandwidth is different[37], as is the case here with the $d_{xy}$ band being narrower than the $d_{xz}/d_{yz}$ bands. When the material is sufficiently close to such an OSMP, the $d_{xy}$ mass renormalization is much stronger than that of other orbitals, as observed in the low temperature state of the FeChs. For the system to be close to this OSMP, it must exhibit strong electron correlations, which from the two theoretical studies is shown to be a combined result of Coulomb interactions $U$ and Hund's coupling $J$.

Among the three FeChs, FTS is undoped, with $n = 6$, while FS/STO and KFS are electron doped, with $n = 6.12$ and $6.15$, as estimated from Fermi surface volume counting. Fig. 6 shows our calculated transition temperatures into the OSMP versus $U$ phase diagrams for systematic dopings from $n = 6$ to $6.15$. The blue shading indicates the $d_{xy}$ quasiparticle spectral weight, $Z_{xy}$, for each doping. When $Z_{xy}$ drops to zero, the system enters the OSMP. This transition boundary is marked by solid blue lines in Fig. 6. Furthermore, we see that for and only for integer filling, $n = 6$ in the calculated doping range, there exists a Mott Insulator (MI) phase (marked by a red line), where all orbitals become Mott localized with sufficiently large $U$. With increasing electron filling, the critical $U$ for the system entering the OSMP at a given temperature shifts to larger values, as can be seen by tracing the blue lines at both 10 K and 240 K, indicating that electron doping brings the system away from the OSMP. This is consistent with the observation that the temperature at which the $d_{xy}$ spectral weight disappears increases from FTS (110 K) to the doped FS/STO (150 K) and KFS (180 K). Here we note that the resistivity for the FeTe$_{1-x}$Se$_x$ family exhibits a weak hump[18] suggestive of a crossover from insulating-like to metallic-like behaviour much like the case of KFS[10], and this crossover temperature scale shifts to lower value with decreasing $x$, consistent with the stronger renormalization values towards the FeTe end[33] and the understanding that the low temperature state of FeTe end sits closer to the OSMP.

**Discussion**

The strong orbital-selective behaviour in the FeCh family is a manifestation of its strong electron correlations, and suggests that they may serve as a bridge between the strongly correlated cuprates as doped Mott insulators and the weaker correlated FePns that are more itinerant. The OSMP of the FeChs can only occur when the overall electron correlations are strong. In that sense, the existence of an OSMP indicates the presence of strong correlations, such that the system is in proximity to a Mott transition. The latter links these materials to the cuprates, where optimal superconductivity develops not too far from a Mott insulating state. The linkage can be made more explicit by the Hunds-coupling-induced suppression of the interorbital coupling in the case of the iron-based materials[19,34,38]. This is relevant to the current discussion on the pairing mechanism of FeSC in both KFS and FS/STO as the lack of hole pockets do not seem to prevent them from superconducting at temperatures comparable to or even higher than FePns, which is unexpected from a weak-coupling Fermi surface nesting picture[3]. In a recent theoretical study[39], it has been shown that under a strong-coupling approach, where the driving force for pairing comes from the close-neighbour exchange interactions, the pairing strength can be comparable in the FeChs and FePns, as they approach a Mott transition by increasing exchange interaction in the former and reducing the renormalized bandwidth in the latter. The observed universal strong correlation in the FeChs and proximity to an OSMP here supports such a scenario.

**Methods**

*Sample Growth*

High quality single crystals of $K_{0.76}Fe_{1.72}Se_2$ and $FeTe_{0.56}Se_{0.44}$ were grown using the flux method[19,40]. Monolayer FeSe films were grown on $SrTiO_3$ using molecular beam epitaxy[17].

*ARPES Measurements*

ARPES measurements were carried out at beamline 5-4 of the Stanford Synchrotron Radiation Lightsource and beamline 10.0.1 of the Advanced Light Source using SCIENTA R4000 electron analyzers. The total energy resolution was set to 10 meV and the angular resolution was 0.3°. Single crystals were cleaved in situ at 10 K for each measurement. The FeSe films were transported to the beamline under vacuum and further annealed before measurements. All measurements were done in ultra high vacuum with a base pressure lower than $4 \times 10^{-11}$ torr.

*Theoretical Calculations*

The theoretical calculations were done using a slave-spin mean-field method on a five-orbital Hubbard model[34], with tight-binding parameters for FTS. Hund's *J* is fixed at 0.6 eV.

**Acknowledgements** ARPES experiments were performed at the Stanford Synchrotron Radiation Lightsource and the Advanced Light Source, which are both operated by the Office of Basic Energy Sciences, U.S. Department of Energy. The Stanford work is supported by the US DOE, Office of Basic Energy Science, Division of Materials Science and Engineering, under award number DE-AC02-76SF00515. The work at Rice is supported by NSF Grant DMR-1309531 and the Robert A. Welch Foundation Grant No. C-1411. The work at Renmin University is supported by the National Science Foundation of China Grant number 11374361, and the Fundamental Research Funds for the Central Universities and the Research Funds of Remnin University of China. The work at Tulane is supported by the NSF under grant DMR-1205469.

**Author Contributions** M.Y., Z.K.L., and Y.Z. performed the ARPES experiments with assistance from M.H. and S.K.M. under the guidance of D.H.L. and Z.H.. R.Y., J.X.Z., and Q.S. performed the theoretical calculations. J.J.L., R.G.M., F.T.S., and W.L. grew the FeSe films. S.C.R., J.H.C., and B.L. grew the KFS single crystals with guidance from I.R.F. and C.W.C.. J.H. grew the FTS single crystals under the guidance of Z.Q.M.. Z.X.S. and D.H.L. provided overall guidance and project coordination. The paper was written by M.Y., D.H.L., R.Y., and Q.S. with contributions from all authors.

**Author Information** The authors declare no competing financial interests. Correspondence and request for materials should be addressed to D. H. Lu (dhlu@slac.stanford.edu) and Z.-X. Shen (zxshen@stanford.edu).

**Fig. 1: Low temperature band structure of iron chalcogenides in comparison to iron pnictide.**

Fermi surfaces measured on (a) FeTe$_{0.56}$Se$_{0.44}$ (FTS), (b) monolayer FeSe film on SrTiO$_3$ (FS/STO), (c) K$_{0.76}$Fe$_{1.72}$Se$_2$ (KFS), and (d) Ba(Fe$_{0.93}$Co$_{0.07}$)$_2$As$_2$ (BFCA), shown in BZ notation corresponding to 2-Fe unit cell (For comparison purposes, we use the *M* point to denote the BZ corner where the electron pockets live for all compounds and LDA, even though for 122 crystal structures, this is the *X* point), with schematic outlines shown in red. (e) Spectral image of FTS along the *Γ-M* high symmetry direction, taken with 22eV (26eV) photons for near the *Γ* (*M*) point. Measurements along the same cut for (f) FS/STO, (g) KFS, and (h) BFCA, with photon energies of 22 eV, 26 eV, and 47.5 eV, respectively. In-plane polarization was odd with respect to the cut for all measurements, (e)-(g) has additional out-of-plane polarization. (i)-(l) Second energy derivatives for the spectral images above. Observable bands are marked with dominant orbital character (red: d$_{xz}$, green: d$_{yz}$, blue: d$_{xy}$).

**Fig. 2: Schematics of the effect of orbital-dependent band renormalizations.** (a) LDA calculations for KFS[41]. (b) Schematic based on (a) with d$_{xy}$ orbital strongly renormalized. (c) Schematic based on (b) by introducing hybridization between d$_{xy}$ band and d$_{xz}$ electron band.

**Fig. 3: Selected spectra from FS/STO.** (a) High resolution spectra of FS/STO showing the presence of two electron bands around *M*. (b) *Γ-M* high symmetry cut of FS/STO taken in the second BZ, where d$_{xy}$ orbital matrix elements are strong. The d$_{xy}$ hole band has a lower band top at *Γ* than d$_{xz}$/d$_{yz}$, hence it crosses and hybridizes with the d$_{xz}$/d$_{yz}$ hole bands, resulting in the apparent sharper curvature near *Γ*.

**Fig. 4: Temperature dependence of the band structure of iron chalcogenides**

(a)-(c) Second energy derivatives of band structure along the $\Gamma$-$M$ cut of FTS, FS/STO, and KFS, same as that of Fig. 1(e)-(g), taken in the low temperature state at 15 K, 9 K, and 10 K, respectively. (d)-(f) Same as that of (a)-(c) but taken in the high temperature state at 120 K, 190 K, and 210 K, respectively. (g) Schematic showing the band structure in the low temperature state with finite spectral weight of $d_{xy}$ orbital, (red: $d_{xz}$, green: $d_{yz}$, blue: $d_{xy}$). (h) Schematic showing the band structure in the high temperature state after $d_{xy}$ orbital completely loses spectral weight.

**Fig. 5: Quantitative analysis of temperature evolution in the iron chalcogenides**

(a)-(c) Raw spectral images of FTS, FS/STO, and KFS taken in the low temperature state. Yellow regions mark the momentum ranges integrated for energy distribution curve (EDC) analysis for each compound. (d) Integrated EDCs in the yellow region of (a) for FTS at selected temperatures, fitted by a Shirley background (gray), a Gaussian for the $d_{xy}$ band (blue), and a Gaussian for the $d_{yz}$ band (green), convolved by the Fermi-Dirac function. (e) Integrated EDCs in the yellow region of (c) for FS/STO at selected temperatures, with a Gaussian background (gray), a Gaussian for the $d_{xy}$ band (blue), and a Gaussian for the $d_{yz}$ band (green). (f) Integrated EDCs in the yellow region of (b) for KFS at selected temperatures, fitted by a Gaussian background (gray), and a Gaussian for the $d_{xy}$ band (blue). Residual spectral weight for the $d_{xy}$ peak is shaded for each temperature for all compounds. Fitted peaks for the lowest temperature are shown for each compound. (g)-(i) Temperature dependence of the fitted areas of the $d_{xy}$ and $d_{yz}$ peaks for FTS,

FS/STO, and KFS. Guides to eye are drawn in gray to show the trends. All curves are normalized by the initial value of the peak area. The error bars in g-i are error bars resulted from the fitting.

**Fig. 6: Calculated phase diagram of the OSMP as a function of temperature, *U*, and electron filling**

Slave-spin mean-field phase diagrams of the five-orbital Hubbard model at systematic electron fillings from $n = 6$ to 6.15. See Methods for calculation details. Blue shading shows the $d_{xy}$ quasiparticle spectral weight, $Z_{xy}$. The OSMP phase boundary is shown by solid blue lines connecting blue squares marking the calculated temperatures where $Z_{xy}$ drops to zero. A Mott Insulator (MI) phase exists for $n = 6$, where all orbitals have zero spectral weight. Its phase boundary is marked by a red line. The temperatures at which $Z_{xy}$ is observed to vanish in FTS, FS/STO, and KFS are marked by magenta dotted lines. From these temperatures, the critical *U*'s for FTS, FS/STO, and KFS can be estimated, which is shown by a yellow guide to the eye strip at fixed *U*. Hund's *J* is fixed at 0.6 eV. Here, electron doping bring the system away from the OSMP, as seen in the increasing critical *U* at fixed temperature (blue lines at 10 K and 240 K) as well as the increasing critical temperature at fixed *U* (magenta dots).

**Figure 1**

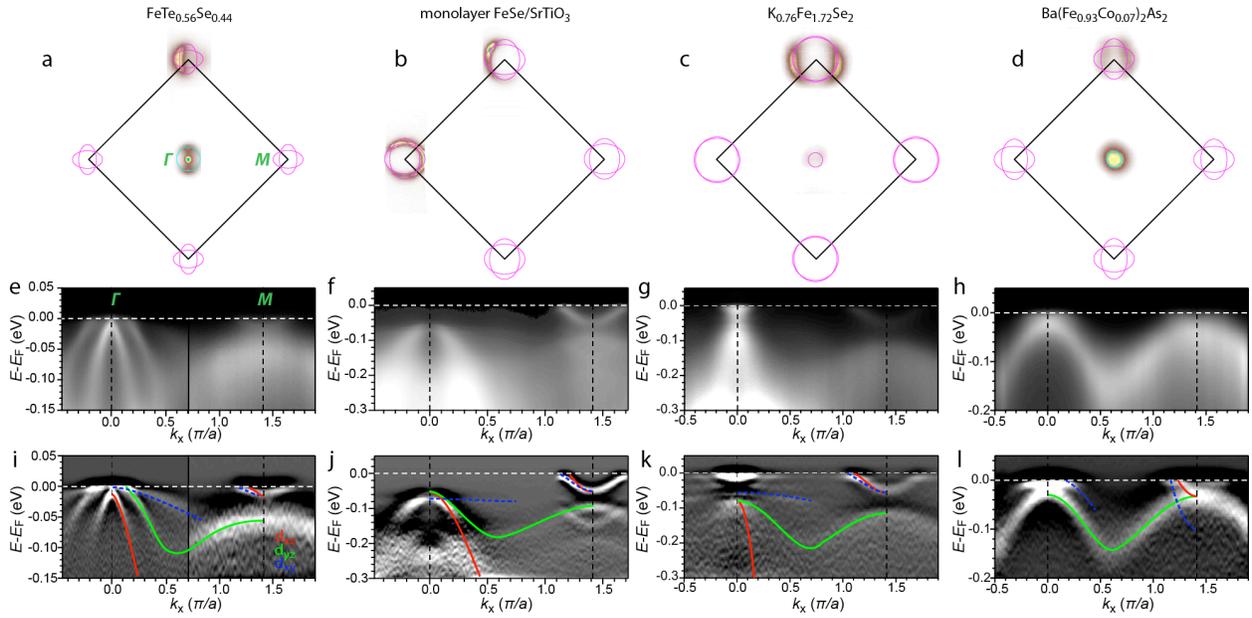

**Figure 2**

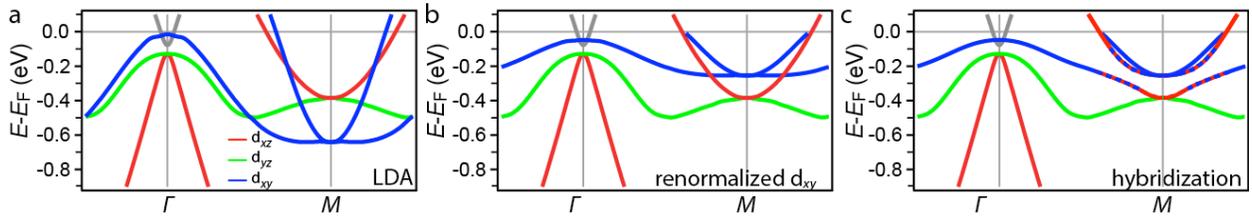

**Figure 3**

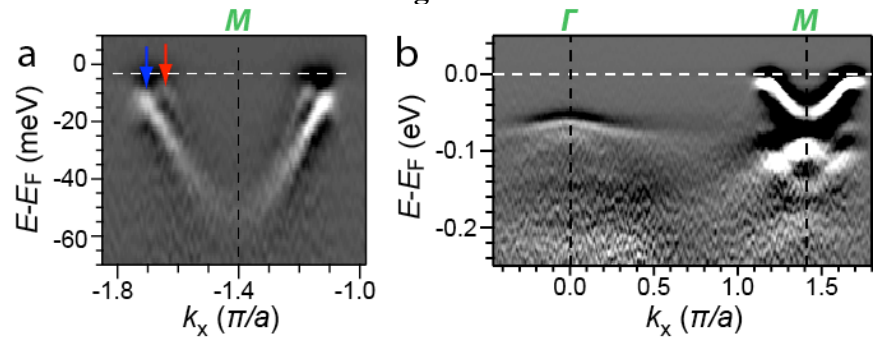

**Figure 4**

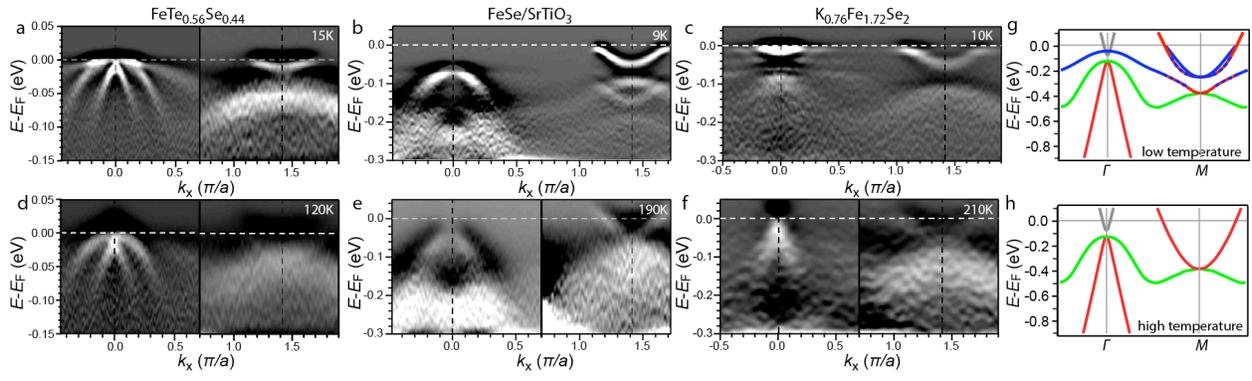

**Figure 5**

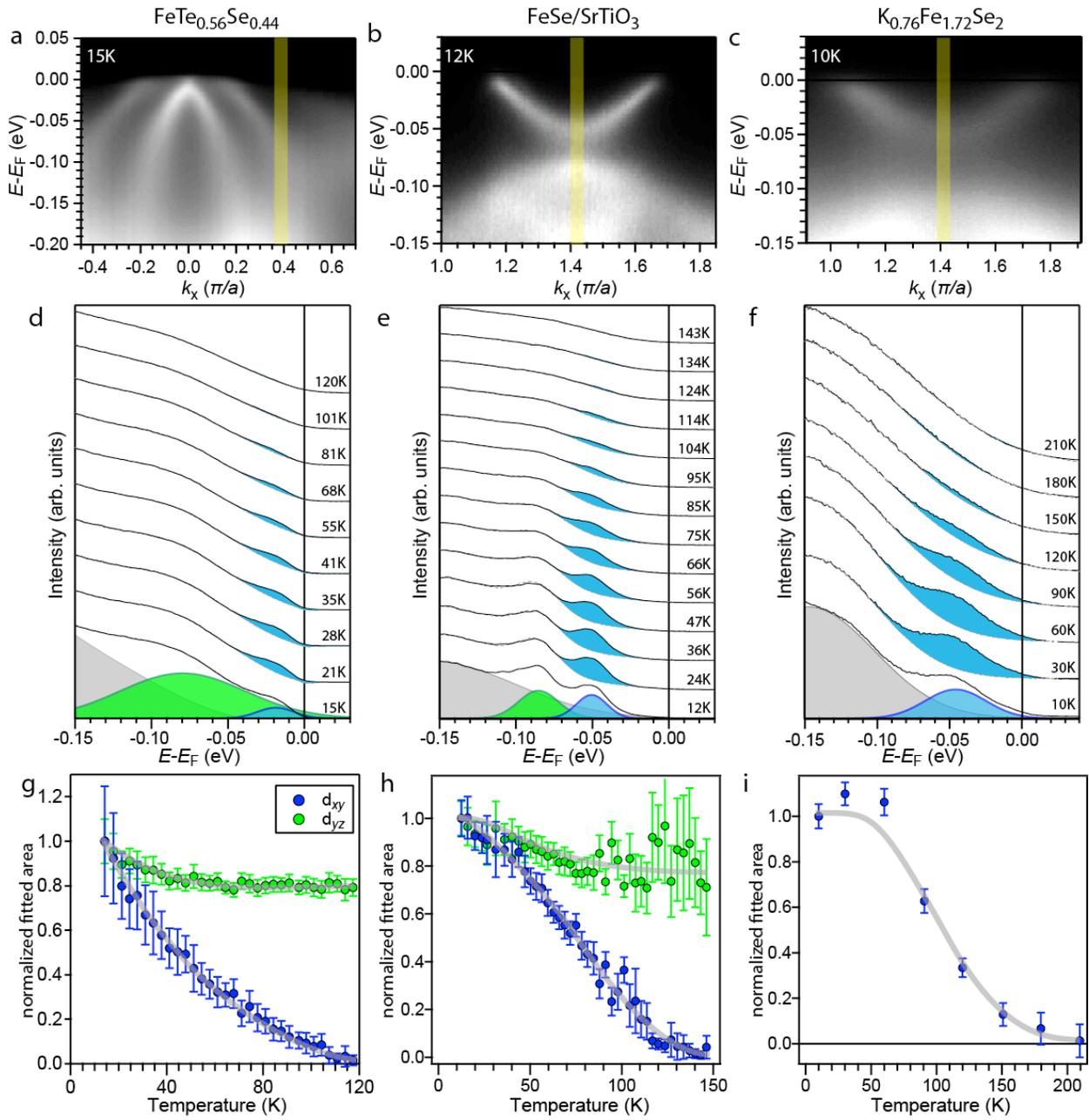

**Figure 6**

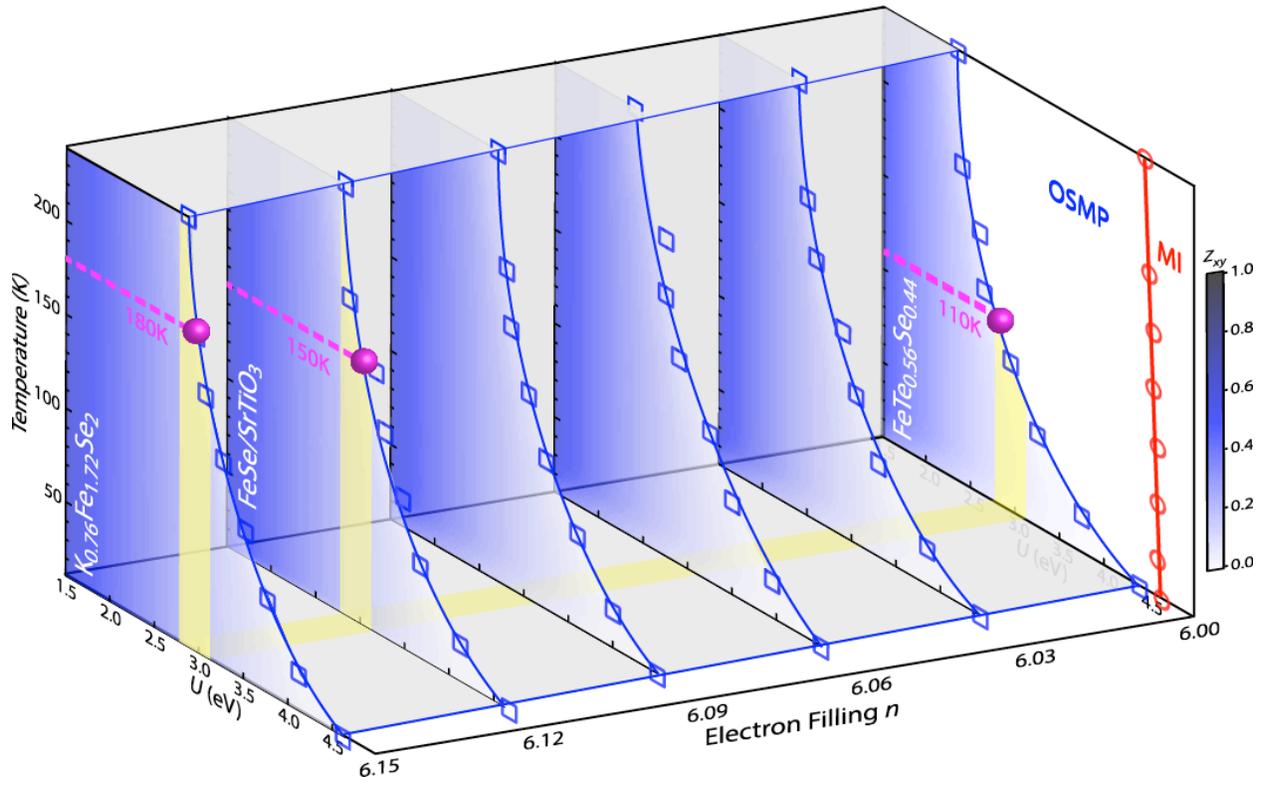